\documentclass{elsart}
\usepackage{graphicx}
\usepackage{amsmath}
\usepackage{amssymb}

\begin{document}

\begin{frontmatter}
\title{Sensitivity of the magnetic state of a spin lattice on itinerant electron orbital phase}

\author[address1]{C. Siegert},
\author[address2]{A. Ghosh},
\author[address1]{M. Pepper},
\author[address1]{I. Farrer},
\author[address1]{D. A. Ritchie},
\author[address1]{D. Anderson}, and
\author[address1]{G. A. C. Jones}

\address[address1]{Cavendish Laboratory, University of Cambridge, J.J. Thomson Avenue, Cambridge CB3 0HE, United Kingdom.}
\address[address2]{Department of Physics, Indian Institute of Science, Bangalore 560 012, India.}

\begin{abstract}
Spatially extended localized spins can interact via indirect exchange interaction through Friedel oscillations in the Fermi sea. In arrays of localized spins such interaction can lead to a magnetically ordered phase. Without external magnetic field such a phase is well understood via a "two-impurity" Kondo model. Here we employ non-equilibrium transport spectroscopy to investigate the role of the orbital phase of conduction electrons on the magnetic state of a spin lattice. We show experimentally, that even tiniest perpendicular magnetic field can influence the magnitude of the inter-spin magnetic exchange.
\end{abstract}

\end{frontmatter}

\section{Introduction}

Ordered arrays of spins, called spin lattices, can form a magnetic state when surrounded by conduction electrons. Such systems are normally characterized by two energy scales. First, the Kondo temperature $T_{\rm K}$, which is the coupling of conduction electrons to localized spins. And second, Ruderman-Kittel-Kasuya-Yosida (RKKY) indirect exchange $J^{\rm RKKY}$, which is the pairwise interaction between localized spins. A mean to investigate the magnetic state of the spin lattice is non-equilibrium transport spectroscopy, where a system with $|J^{\rm RKKY}| = 0$ shows a resonance in differential conductance ${\rm d}I/{\rm d}V$ at zero source-drain bias ($V_{\rm SD}$) \cite{kouwenhoven2001,ghosh2005}, which we call type-I zero bias anomaly (ZBA). With interacting spins $|J^{\rm RKKY}| > 0$, the resonance is suppressed at $E_{\rm F}$ for $k_{\rm B} T < |J^{\rm RKKY}|$, and $|eV_{\rm SD}| < |J^{\rm RKKY}|$, resulting in a split resonance, which we call type-II ZBA \cite{heersche2006,wiel2002,ghosh2004,siegert2007}. In two dimensions electron wavevector $k_{\rm F} = \sqrt{2 \pi n_{\rm 2D}}$ and interspin distance $R$ tunes the exchange interaction parameter $J^{\rm RKKY}(k_{\rm F},R) \propto {\rm cos}(2 k_{\rm F} R)/R^2$ for $k_{\rm F}R >> 1$ \cite{glazman2004,Beal-Monod1987}. Such a tunig has recently been shown in an $intrinsic$ lattice of spins, that was reported to exist in high mobility modulation doped two dimensional electron gases (2DEG) \cite{siegert2007}. Theoretically, spin lattices are often analysed in a "two-impurity" Kondo model \cite{cox1996}, where the effect of perpendicular magnetic field, and its effect on the magnetic state via the orbital phase of the electrons that couple the spins, is not included. However, even in two spin systems, $J^{\rm RKKY}$ is strongly influenced by the nature and extent of confinement of the intervening Fermi sea, and the orbital phase of itinerant electrons \cite{craig2004,simon2005,vavilov2005,utsumi2004}. We investigate such influence by application of a small transverse magnetic field on a quasi-regular two-dimensional array of localized spins, embedded within a sea of conduction electrons. We show that the magnetic state is extremely sensitive to perpendicular field $B_\bot$, and the orbital phase of the itinerant electrons.

\section{Experiment}
\begin{figure}\label{Fig1}\centering
\includegraphics[width=7cm]{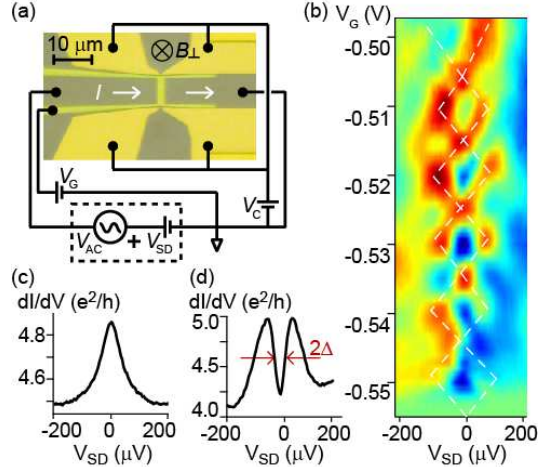}\caption{(a) Optical microscope image of a typical $2 \times 8$ $\mu {\rm m}^2$ device, showing the electrical setup. Negative voltage on $V_{\rm C}$ confines electrons in 8 $\mu {\rm m}$ wide mesa. $V_{\rm G}$, and $V_{\rm SD}$ are used to vary electron density, and source-drain voltage respectively, in active area. (b) On variation of the full gate $V_{\rm G}$ a modulation of ${\rm d}I/{\rm d}V(V_{\rm SD})$ is observed, alternating between type-I and type-II ZBA. (red/blue: high/low conductance. White line: guide to the eye) (c) Typical type-I ZBA, showing a single peak at ${\rm d}I/{\rm d}V(V_{\rm SD} = 0)$. (d) ${\rm d}I/{\rm d}V(V_{\rm SD})$ of a type-II ZBA, with half width of half depth of the dip defined as $\Delta$.}
\end{figure}

We probe the effect of perpendicular magnetic field on spin ordering in systems with an intrinsic two-dimensional spin lattice by means of non-equilibrium magnetoconductance spectroscopy. We use Si-monolayer-doped Al$_{0.33}$Ga$_{0.67}$As/GaAs heterostructures. Using a 60-80 nm spacer between 2DEG and Si results in as-grown electron density of $n_{\rm 2D} = 1 - 1.5 \times 10^{11}$ cm$^{-2}$ with mobilities $\mu = 1 - 3 \times 10^6$ $\frac{{\rm cm}^2}{\rm Vs}$. We thermally deposit non-magnetic Ti/Au gates on the surface, which allows us to locally tune $n_{\rm 2D}$ of the 2DEG, 300 nm below the surface. Fig.~1(a) shows a typical device, and electrical connections. Voltage $V_{\rm C}$ is used to deplete electrons by application of up to -1 Volt, which defines an 8 $\mu {\rm m}$ wide mesa electrostatically. $V_{\rm C}$ is kept fixed for the duration of the experiment. Alternatively an etched mesa with same width can be used to create lateral confinement. $V_{\rm G}$ is biased such that the devices operate in the range $n_{\rm 2D} = 0.5-2 \times 10^{10}$ cm$^{-2}$ in a $2 \times 8$ $\mu {\rm m}^2$ region. Constant voltage two probe combined AC+DC technique is used with $e V_{\rm AC} \ll k_{\rm B}T$ for all $T$, and source-drain bias $V_{\rm SD}$. Fig.~1(b) shows a typical non-equlibrium spectroscopy image of ${\rm d}I/{\rm d}V (V_{\rm SD},V_{\rm G})$ at $T = 75$ mK and zero external field. Individual ${\rm d}I/{\rm d}V (V_{\rm SD})$ show either a type-I ZBA (as shown in fig.~1(c)), or a type-II ZBA (as shown in fig.~1(d)). For the type-II ZBA, the half width at half depth is defined as $\Delta$ and is proportional to the exchange interaction parameter: $\Delta \propto |J^{\rm RKKY}|$ \cite{siegert2007,ghosh2005}.

\section{Results and Discussion}
\begin{figure}\label{Fig2}\centering
\includegraphics[width=7cm]{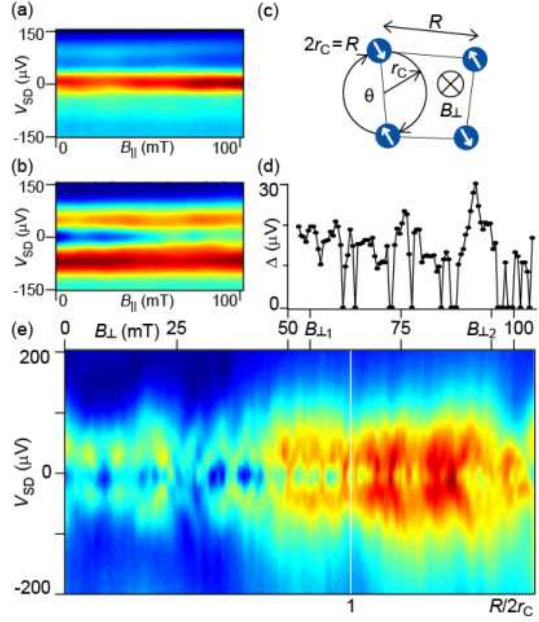}\caption{(a) Parallel field dependence of a type-I ZBA showing neither a modulation of $\Delta$ nor of $G$ (b) Parallel field dependence of type-II ZBA also showing no modulation. (c) In perpendicular magnetic field $B_\bot$ spin interaction is maximized for commensurable fields, where $2r_{\rm C} = R$, and electrons acquire an additional phase $\theta$. (d) $\Delta$ vs. $B_\bot$ for fields between 50 mT and 100 mT. (e) Non-equilibrium conductance evolution of typical type-II ZBA over low perpendicular magnetic field (red: high / blue: low ${\rm d}I/{\rm d}V$). Around the commensurability condition $2r_{\rm C} = R$ (white line), the overall ZBA magnitude increases over a range between 55 and 95 mT.}
\end{figure}

In figures~2(a), and 2(b) we show the evolution of type-I and type-II ZBA in low parallel magnetic field. Both ZBA's are not affected by parallel magnetic field in that range, and only show a splitting at higher fields \cite{siegert2007}. In contrary, perpendicular magnetic field $B_\bot$ forces conduction electrons on paths with cyclotron radii $r_{\rm C} = \frac{\hbar k_{\rm F}}{e B_\bot}$, and electrons aquire additional phase $\theta$, see fig.~2(c). When $R = 2 r_{\rm C}$, electrons from one localized spin are focussed on the next, and vice versa, which can be seen as an enhancement of overall ZBA magnitude, and of fluctuations in $\Delta$ \cite{usaj2005}. We show such commensurability behaviour in a surface plot of the non-equilibrium conductance over $B_\bot$, up to 105 mT for a typical type-II ZBA, see fig.~2(e). The commensurability condition is indicated by the white line in fig.~2(e) at the field $B_\bot^{\rm C} \approx 65$ mT for interspin distance $R = 630$ nm. The intrinsic interspin distance $R$ is evaluated from Aharonov-Bohm (AB) oscillations in linear ($V_{\rm SD} = 0$) magnetoconductance at low $T$ \cite{siegert2007}. In fig.~2(d) we show the respective $\Delta (B_\bot)$, and note that above 50 mT we observe strong and irregular fluctuations in $\Delta$, which is substantially different from the low field part.

In fig.~3(a) we show that $\Delta$ is periodically suppressed with a period $\Delta B \approx 9 - 10$ mT at fields $B_\bot \ll B_\bot^{\rm C}$. In the top of the image $\Delta (B_\bot)$ is shown. Since $\Delta \propto J^{\rm RKKY}$, the magnetic state of the spin lattice is directly modulated by perpendicular magnetic field. Theoretically this has been simulated for the two-impurity case, where models link magnetic flux and $J^{\rm RKKY}$ in $B_\bot$, and predict a modulation according to \cite{utsumi2004,nitta1999,schuster1994}:
\begin{equation}\label{deltaphi}
J^{\rm RKKY} (B_\bot \neq 0) \sim (\frac{1}{2} + \frac{1}{2} {\rm cos}(\theta)) \cdot J^{\rm RKKY} (B_\bot = 0),
\end{equation}
with $\theta(B_\bot)$ being the orbital phase shift in $B_\bot$ from the electron trajectories on their paths from one spin to the next. The periods of AB and $\Delta$ oscillations can be compared. Both, AB oscillations, as well as eq. (\ref{deltaphi}) have a $B_\bot$-periodicity of $\Delta \theta = 2 \pi$, and thus the periods are similar. Please note that AB oscillations can not influence $\Delta$, and thus non-equilibrium spectroscopy can be used to distinguish them from $\Delta$ oscillations. 

\begin{figure}\label{Fig3}\centering
\includegraphics[width=7cm]{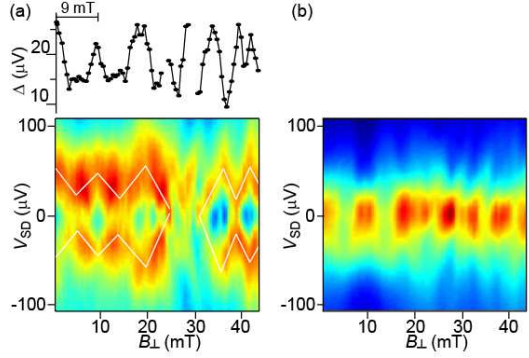}\caption{ (a) Top: Variation of $\Delta$ over perpendicular magnetic field, showing a modulation with rough periodicity of $\Delta B \approx 9 - 10$ mT. Bottom: Respective non-equilibrium traces in $V_{\rm SD}$ - $B_\bot$ plane. (b) Low field non-equlibrium magnetoconductance of a type-I ZBA showing modulation of $G$, but not of $\Delta$.}
\end{figure}

An interesting consequence of eq. (\ref{deltaphi}) is, that small fields ($B_\bot < B_\bot^{\rm C}$) can only decrease $J^{\rm RKKY}$, but not increase it from the zero field value. A type-I ZBA with $J^{\rm RKKY} = 0$ would then show AB oscillations, but have no $\Delta$ modulation over $B_\bot$. We do indeed observe this, and show in fig.~3(b) a typical surface plot of non-equilibrium low field magnetoconductance of a type-I ZBA, where the linear conductance $G(B_\bot)$ is modulated without a modulation of $\Delta$. Thus, low perpendicular field can not change the magnetic state of a spin lattice with $J^{\rm RKKY} = 0$.

Please note that in our system the background disorder plays a critical role in the arrangement of the localized spins. While the origin of the localized spins is not fully understood yet, periodic tunability of $\Delta$ seems to be only possible in a certain window of disorder. Unintentional disorder can lead to additional scattering of electron trajectories in $B_\bot$, which suppresses clear observation of $B_\bot$-tuning of $\Delta$. Even in the same device, about 25\% of electron densities displayed no periodic modulation of $\Delta$ with $B_\bot$, even though the oscillation in $\Delta$ as a function of $k_{\rm F}$ at $B_\bot = 0$ was clearly observable.

\section{Summary}

At low temperatures non-equilibrium conductance spectroscopy in low perpendicular magnetic fields is used to analyse the magnetic state of a spin lattice. We show that even tiny fields modulate the indirect exchange interaction $J^{\rm RKKY}$ between localized spins, directly influencing the magnetic state of the spin lattice.

\section{Acknowledgement}

This project was supported by EPSRC. CS acknowledges the support of Cambridge European Trust, EPSRC, and Gottlieb Daimler and Karl Benz Foundation.


\end{document}